\documentclass[12pt]{iopart}
\usepackage{iopams}  
\usepackage{color}
\usepackage[dvipsnames]{xcolor} % Before pstricks: permits OliveGreen, etc.
\usepackage{tikz}
\usetikzlibrary{decorations.pathmorphing,patterns} %% For coil decoration
\usepackage{psfrag}
\usepackage{graphicx}
\usepackage{stackrel}
\usepackage{amsfonts}
\usepackage{bm}          %% For bold math symbols
\usepackage{amssymb}
\usepackage{perpage}
\MakePerPage{footnote}

\usepackage{geometry}                % See geometry.pdf to learn the layout options. There are lots.
\usepackage{graphicx}
\usepackage{epstopdf}
\begin{document}

\title{On the growth constant for square-lattice self-avoiding walks}
\author{Jesper Lykke Jacobsen$^{1,2,3}$, Christian R. Scullard$^{4}$ and Anthony J. Guttmann$^{5}$}
\address{${}^1$LPTENS, \'Ecole Normale Sup\'erieure -- PSL Research University, 24 rue Lhomond, F-75231
Paris Cedex 05, France}
\address{${}^2$Sorbonne Universit\'es, UPMC Universit\'e Paris 6, CNRS UMR 8549, F-75005 Paris, France} 
\address{${}^3$Institut de Physique Th\'eorique, CEA Saclay, F-91191 Gif-sur-Yvette, France}
\address{${}^4$Lawrence Livermore National Laboratory, Livermore CA 94550, USA}
\address{${}^5$ARC Centre of Excellence for Mathematics and Statistics of Complex Systems, Department of Mathematics and Statistics,
The University of Melbourne, Victoria 3010, Australia}

%\date{}                                           % Activate to display a given date or no date

%\maketitle

\begin{abstract}
The growth constant for two-dimensional self-avoiding walks on the honeycomb lattice was conjectured by Nienhuis in 1982, and since that time the corresponding results for the square and triangular lattices have been sought. For the square lattice, a possible conjecture was advanced by one of us (AJG) more than 20 years ago, based on the six significant digit estimate available at the time. This estimate has improved by a further six digits over the intervening decades, and the conjectured value continued to agree with the increasingly precise estimates. 
We discuss the three most successful methods for estimating the growth constant, including the most recently developed Topological Transfer-Matrix method, due to another of us (JLJ). We show this to be the most computationally efficient of the three methods, and by parallelising the algorithm we have estimated the growth constant significantly more precisely, incidentally ruling out the conjecture, which fails in the 12th digit. Our new estimate of the growth constant is
$$\mu(\mathrm{square}) = 2.63815853032790\, (3).$$
\end{abstract}

\section{Introduction}

In 1982 Nienhuis \cite{Nienhuis82} conjectured that the growth constant for self-avoiding walks (SAWs) on the hexagonal lattice was the algebraic number
$\mu_H=\sqrt{2+\sqrt{2}},$ a result finally proved thirty years later by Duminil-Copin and Smirnov \cite{DuminilSmirnov12}. Note that $\mu_H$ satisfies
$\mu_H^2-4\mu_H+2=0.$ Given that the corresponding quantity---the inverse critical temperature---for the two-dimensional Ising model on all
regular two-dimensional lattices is also algebraic, it appears likely, or at least plausible, that the growth constant for SAWs on the square and
triangular lattices is also algebraic. This prompted one of us (AJG) in the 1980s to take the best currently available estimates of the growth
constants for these lattices, and to search for an integer polynomial with root equal to these estimates.  Based on 6-digit accuracy in the estimate
of $\mu$ for the square lattice at the time, it was found that the polynomial
\begin{equation}
 13t^4-7t^2-581
 \label{polynomial}
\end{equation}
has a real root $t=2.6381585303417408684303\cdots$
which agreed with the best current estimate of $\mu.$ The corresponding conjecture for the radius of convergence of the SAW generating function is thus
\begin{equation}
  x_{\rm c}^{\rm conj} = 1/\mu = 0.37905227775317290937028\cdots .
  \label{the-conjecture}
\end{equation}

Over the intervening years, indeed decades, as the estimate of the value of $\mu$ became increasingly more precise, this polynomial root continued to
satisfy the current best estimate. For example, in 2001, Guttmann and Conway \cite{GuttmannConway01} quoted $\mu=2.638158534(4)$ as the best current estimate,
based on an analysis of the self-avoiding polygon (SAP) series for polygons up to perimeter 90.\footnote{It is well-known that the SAP and SAW series
have the same growth constant \cite{Hammersley61}.} Eleven years later, Clisby and Jensen \cite{ClisbyJensen12} improved the algorithm for SAP enumeration
to perimeter 130. Analysing this extended series, they estimated $\mu=2.63815853035(2).$ So the original quadratic mnemonic based
on a 6-digit estimate of $\mu$ is seen to hold for 12 digits. Despite this encouraging outcome, one problem with the original quartic (\ref{polynomial}) is that while it has a second
zero at $-2.6381585303417408684303\cdots$, corresponding to the ``anti-ferromagnetic" singularity, known to be present in the SAW generating function,
there is also a conjugate pair of singularities on the imaginary axis, and numerical analysis of both the SAW generating function and the SAP generating
function has revealed no presence of such a singularity.

Until now, extrapolation of SAP series has been the most precise method for estimating $\mu$. A competing method, based on an adaptation of an identity
found by Duminil-Copin and Smirnov \cite{DuminilSmirnov12} was developed by Beaton, Guttmann and Jensen \cite{BeatonGuttmannJensen12}, but has not been pushed to its full potential.
Building on earlier work by the other two authors (JLJ and CRS) on a topologically weighted graph polynomial for the Potts model
\cite{JacobsenScullard12,JacobsenScullard13,Jacobsen14}, one of us (JLJ) recently proposed a third method for estimating the critical point of O($N$) loop models
\cite{Jacobsen15}. It is based on equating the eigenvalues of the transfer
matrix for a semi-infinite cylinder of circumference $n$ in two distinct topological sectors, and preliminary results up to $n=19$ were already given in \cite{Jacobsen15}.
We here present a parallel implementation of this topological transfer matrix method, which permits us to attain $n=21$. Analysing these data, the resulting
precision is shown to be superior to that currently obtained by the other two methods. In particular, we shall show that the conjecture (\ref{the-conjecture})
is too low by about $2 \cdot 10^{-12}$.

Below we discuss these three methods, while paying special attention to their computational complexity and convergence properties. Since the computational
effort that has been spent on either method is not identical (and cannot easily be compared), it is of interest to determine which method has the largest potential
for future improvements. We defer the outcome of this comparison to the concluding section.

\section{Series generation and analysis}
\label{sec:series}

The coefficients of the polygon generating function are obtained from transfer matrices on finite lattices (rectangles, in the case of the square lattice), giving rise to the name of this approach as the Finite Lattice Method (FLM). We give a very brief description of the method, following the development given in \cite{EntingJensen09}, which gives much more detail. It turns out that analysing the polygon generating function gives much greater precision than analysing the SAW generating function, as a lattice of any given finite size contains polygons of perimeter approximately twice that of the corresponding walks contained in the same finite lattice. Put another way, the FLM is particularly well-suited to enumerating polygons.

Enting \cite{Enting80} was the first to use transfer matrix techniques to
enumerate self-avoiding polygons. The next qualitative advance was the use of {\em pruning} by
Jensen and Guttmann \cite{JensenGuttmann99} to produce an exponentially faster
algorithm. Jensen \cite{Jensen03} has also implemented efficient parallel
versions of the algorithm, and more recently still Clisby and Jensen \cite{ClisbyJensen12} have developed a more efficient algorithm, allowing polygons to be counted up to perimeter 130 steps.

The generating function
for the number of SAPs per vertex of the infinite lattice is obtained by combining the
contributions from finite sub-lattices. On the square lattice one uses
 rectangles $w$ cells wide and $\ell$ cells long.  Due to the lattice symmetry
one need only consider rectangles with $\ell \geq w$, and one counts the number of
polygons of length exactly $\ell$ and width exactly $w$,
that is polygons which touch all four sides of the rectangle.

In applying the transfer matrix technique to the enumeration of polygons
we regard them as sets of edges on the finite lattice with the properties:
\begin{itemize}
\item[(1)] A weight $x$ is associated with each occupied edge.
\item[(2)] All vertices are of degree 0 or 2.
\item[(3)] Apart from isolated sites, the graph has a single connected
component.
\item[(4)] Each graph spans the
rectangle from left to right and from bottom to top.
\item[(5)] Each column is built up by adding a single lattice cell at a time.
\end{itemize}

In the original application \cite{Enting80}, valid polygons were
required to span the enclosing rectangle only in the lengthwise direction. It turns out to be more efficient to require the polygon to span the rectangle in both directions.

\begin{figure}
\begin{center}
\includegraphics[width=12cm]{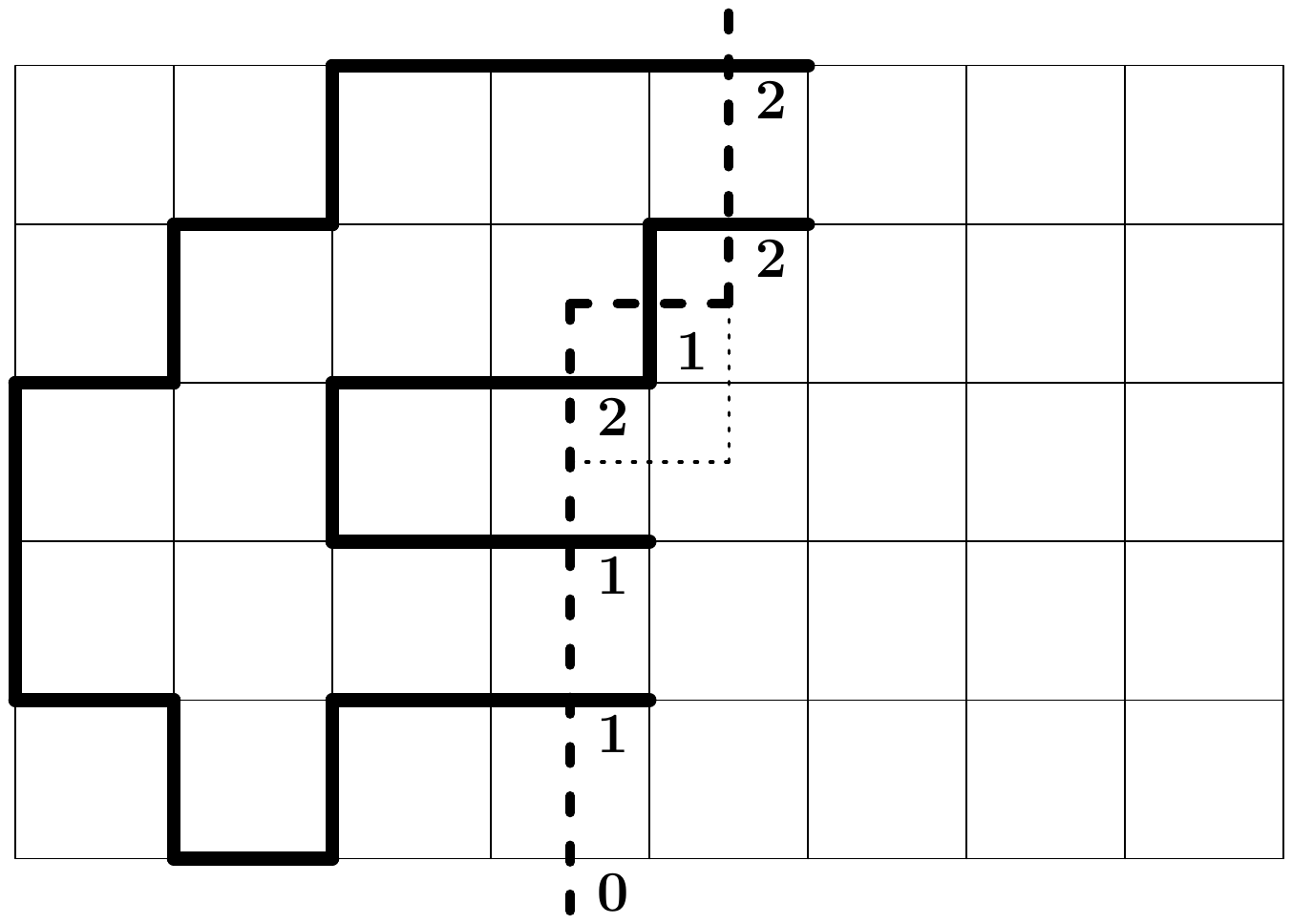}
\end{center}
\caption[Transfer matrix for self-avoiding polygons on the square lattice]
{\label{fig:saptransfer}
A snapshot of the boundary line (dashed line) during the transfer matrix
calculation on the square lattice. Polygons are enumerated by successive
moves of the kink in the boundary line, as exemplified by the position given
by the dotted line, so that one vertex at a time is added to the rectangle.
To the left of the boundary line we have drawn an example of a
partially completed polygon. The numbers along the boundary line
is the encoding of the edge states of the loops intersected by the boundary line.}
\end{figure}

The transfer matrix technique  involves drawing a boundary line through the
rectangle intersecting a set of up to $w+2$ edges. Polygons in a given rectangle
are enumerated by moving the boundary line so as to add one vertex or lattice cell
at a time, as shown in Figure \ref{fig:saptransfer}. The rectangle is built up
column by column with each column built up vertex by vertex.  As the
boundary line moves through the rectangle it intersects partially completed polygons consisting of disjoint
loops. Eventually all the loops must be connected to form a single polygon. For each
configuration  of occupied or empty edges along the intersection one maintains a
(perimeter) generating function for open loops to the left of the line cutting the
intersection in that particular pattern. The updating of the generating
functions depends primarily on the states of the two edges at the kink
in the boundary line prior to the move (called
{\em kink edges}). As the boundary line is moved the two new edges intersected
by the boundary line can be either empty or occupied.  

The constraints listed above must all be satisfied. Constraints 1, 2, 4 and 5 are relatively easy to satisfy, but satisfying constraint 3 requires considerable care. For details see \cite{EntingJensen09}.

One can get further savings in time and memory usage by {\em pruning}. This procedure
involves discarding most of the possible
configurations for large $w$ because they only contribute to polygons of
length greater than $4w_{\rm max}+2$. Briefly this works as follows.
Firstly, for each configuration one keeps track of the current minimum
number of steps $n_{\rm cur}$ already inserted to the left of the boundary
line in order to build up that particular configuration. Secondly, one
calculates the minimum number of additional steps $n_{\rm add}$ required to
produce a valid polygon. There are three contributions, namely the number
of steps required to close the polygon, the number of steps needed (if any)
to ensure that the polygon touches both the lower and upper border, and
finally the number of steps needed (if any) to extend at least $w$ edges
in the length-wise direction (one only needs rectangles
with $\ell \geq w$). If the sum $n_{\rm cur}+n_{\rm add} > 4w_{\rm max}+2,$ 
 the partial generating function for that configuration can be discarded because it will not
contribute to the polygon count, up to the perimeter lengths sought. For instance, polygons spanning a rectangle with a width
close to $w_{\rm max}$ have to be almost convex, so very convoluted
polygons are not possible. Thus configurations with
many loop ends make no contribution at perimeter
length $\leq 4w_{\rm max}+2$.

Symmetries of the underlying lattice can be used to further reduce the number of
configurations  needed to be retained.
There is the basic symmetry of the square lattice allowing one to reduce the computational complexity, since only
rectangles with $\ell \geq w$  need be considered. Moreover, after
a column has been completed (and the boundary line is completely
vertical) configurations are symmetric with respect to reflections.
 Given the symmetry of the square lattice
it is clear that quite a number of partially completed polygons must have a matching symmetric polygon.
 So their generating functions must
be identical and one can discard one while multiplying
the other generating function by 2.

The time required to obtain the number of polygons on $w\times \ell$ rectangles grows
exponentially with $w$. Time and memory requirements are
basically proportional to the maximum number of distinct configurations along the
boundary line. When there is no kink in the intersection (a column has just been
completed) this number, $N_{\rm conf}(w)$ can be calculated exactly. Each
boundary line configuration is encoded by
`0's and an equal number of `1's and `2's with the latter forming
a perfectly balanced parenthesis system.
This corresponds to a Motzkin path.
The number of Motzkin paths $M_n$ with $n$ steps is well known
from the generating  function
\begin{equation}
 {\mathcal M}(t) =  [1-t-\sqrt{(1+t)(1-3t)}]/2t^2 \,,
\end{equation}
and the coefficient of $t^n$ grows like $3^n.$

When the boundary line has a kink the number of configurations exceeds $N_{\rm conf}(w)$
but clearly is less than $N_{\rm conf}(w+1)$. Asymptotically
$N_{\rm conf}(w)$ grows like $3^w$ (up to a power of $w$). So the same is true for the maximal
number of boundary line configurations and hence for the computational complexity of the
algorithm. Note that the total number of SAPs grows like $\mu^{n}$
(where $\mu \simeq 2.638$), while the complexity of the
transfer matrix algorithm grows as $3^{n/4}$. Since $\sqrt[4]{3}\simeq 1.316$
it is clear that even the basic algorithm
without pruning leads to a very substantial exponential improvement over direct enumeration.
Pruning results in a further exponential improvement to the algorithm. In this case
 we find a growth constant of only
$\lambda \approx 2.06^{1/4}\simeq 1.198\ldots$.

In 2011, Clisby and Jensen \cite{ClisbyJensen12} made a further significant improvement by pointing out that the TM algorithms described above all  keep track of the way
partially constructed SAPs are connected to the left of a line bisecting the given rectangles. The new approach keeps track of how partially constructed SAP must connect to the right
of the boundary line.%
\footnote{An alternative description (not contained in the original work \cite{ClisbyJensen12}) of this algorithm is to say that the configuration is built up by multiplications of the {\em transpose} of the transfer matrix.}
The major gain is that it is now straightforward to calculate the number of additional bonds required to complete a given partial SAP. This 
results in a substantially faster algorithm. The draw-back is that some updating rules become much more complicated. The result is an improvement of 15\% in memory usage and a reduction of some 30\% in time.

Using this algorithm, Clisby and Jensen obtained polygons to perimeter 130. Analysing this series by the method of differential approximants \cite{Guttmann89}, they estimated $\mu = 2.63815853035(2),$ or
\begin{equation}
  x_{\rm c} = 0.379052277752 (3) \,.
\end{equation}
This is the most precise numerical value of $x_{\rm c}$ published prior to the present paper.
One sees that it agrees with the value (\ref{the-conjecture}) of the mnemonic polynomial (\ref{polynomial}), with uncertainty confined to the 12th digit.

\section{Adaptation of Duminil-Copin and Smirnov's identity}

\begin{figure}
\includegraphics[width=4in,angle=270]{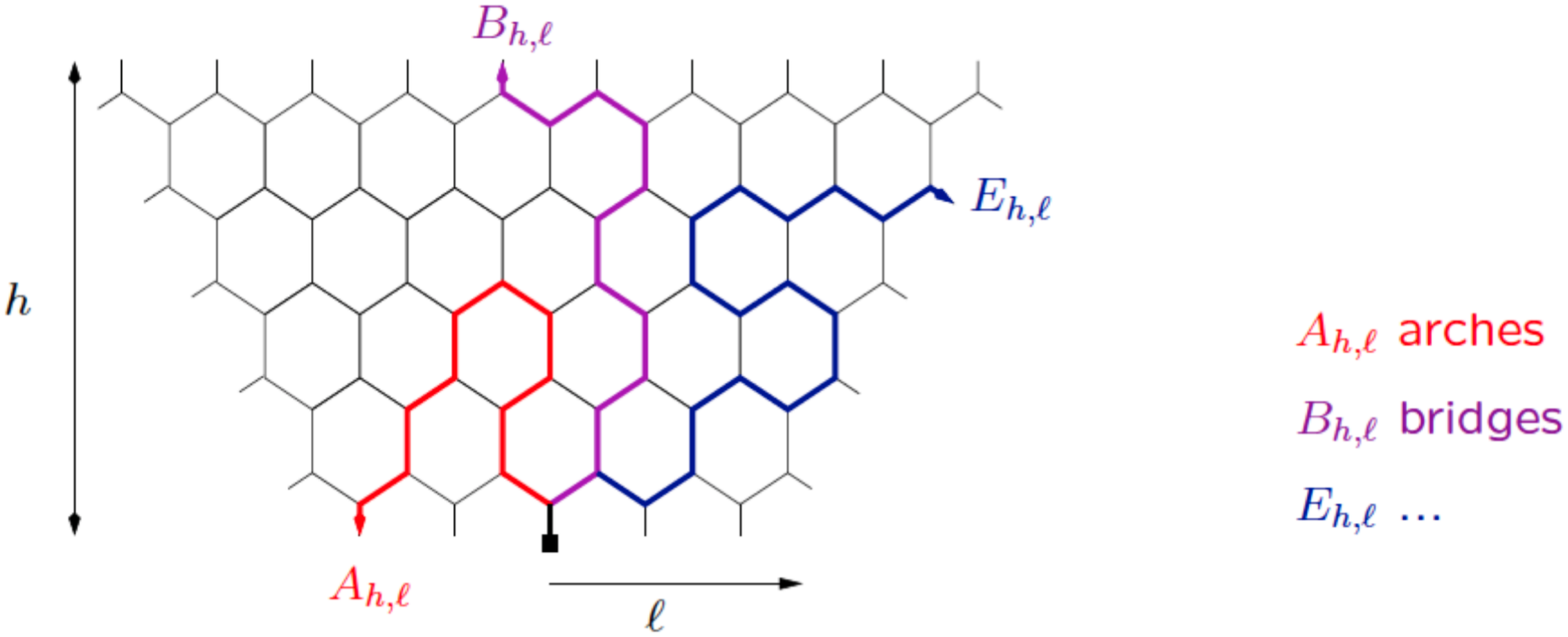}
\caption{Different types of self-avoiding walks in a wedge geometry.
%\textcolor{red}{\bf In this figure, $E_{h,\ell}$ should be called ``end-walks''.}}
}
\label{fig:DCS}
\end{figure}

Duminil-Copin and Smirnov \cite{DuminilSmirnov12} proved a key identity that links three generating functions for a finite-sized wedge-shaped
piece of the hexagonal lattice, as shown in Figure~\ref{fig:DCS}. (Of course, in this situation the generating functions are polynomials). For technical reasons we consider walks starting and ending at the mid-point of bonds, rather than the usual convention of starting and ending at a vertex. The wedge-shaped sector is of length $2\ell$ at its bottom edge, and of height (width) $h.$ All walks start at the mid-point of the bottom edge. We define {\em arches} as walks that also end at the bottom edge, with generating function $A_{h,\ell}(x),$ while {\em bridges} end at the top edge, and have generating function $B_{h,\ell}(x),$ whereas {\em end-walks}, with generating function $E_{h,\ell}(x),$ end at the left or right (sloping) side of the wedge-shaped region. Of course some walks also end in the interior of the sector. However the power of the identity is that, for a particular choice of $x$, namely $x=x_{\rm c}=1/\mu_H$, the walks ending internally make no contribution. The identity is
\begin{equation}
 \cos \left ( \frac{3\pi}{8} \right )A_{h,\ell}(x_c)+B_{h,\ell}(x_{\rm c})+\cos \left ( \frac{\pi}{4} \right )E_{h,\ell}(x_{\rm c})= 1 \,.
\end{equation}
What is remarkable is that this identity connects properties of the finite lattice, the three polynomial generating functions, with a bulk property, the radius of convergence of the generating functions on an infinite lattice.

If one now lets $\ell \to \infty,$ so that the wedge-shaped region becomes a strip of width $h,$ it is easy to show that $E_{h,\ell}(x_{\rm c}) \to 0,$ and the identity simplifies to
\begin{equation}
 \cos \left ( \frac{3\pi}{8} \right )A_{h}(x_{\rm c})+B_{h}(x_{\rm c})= 1 \,.
 \label{DCS-identity}
\end{equation}
This holds for all values of $h$. We also know, from SLE theory and from numerical work, that $B_h(x_{\rm c}) \sim b_1 \cdot h^{-1/4}.$ From this it follows that
$\cos \left ( \frac{3\pi}{8} \right ) \cdot A_h(x_{\rm c}) \sim 1-b_1 \cdot h^{-1/4},$ so among other things, $\lim_{h \to \infty} A_h(x_{\rm c})  = \sec \left ( \frac{3\pi}{8} \right )$.
Now the $h$ dependence just given of
$A_h(x_{\rm c})$ and $B_h(x_{\rm c})$ must be the first term of an asymptotic series. In particular, if we write
\begin{eqnarray}
 A_h(x_{\rm {\rm c}}) &=& a_0+ \sum_{i \ge 1} \frac{a_i}{h^{\alpha_i}} \,, \nonumber \\
 B_h(x_{\rm c}) &=& \sum_{i \ge 1} \frac{b_i}{h^{\beta_i}} \,,
\end{eqnarray}
it follows, for the hexagonal lattice, that
\begin{equation}
 \alpha_i = \beta_i \,, \quad {\rm and} \quad \cos(3\pi/8) a_i +b_i =0 \,,
\end{equation}
{\em for all $i>0$}. This seems quite remarkable, and is a very special property of these generating functions on the hexagonal lattice only.

\begin{figure}
\begin{center}
\includegraphics[width=3in]{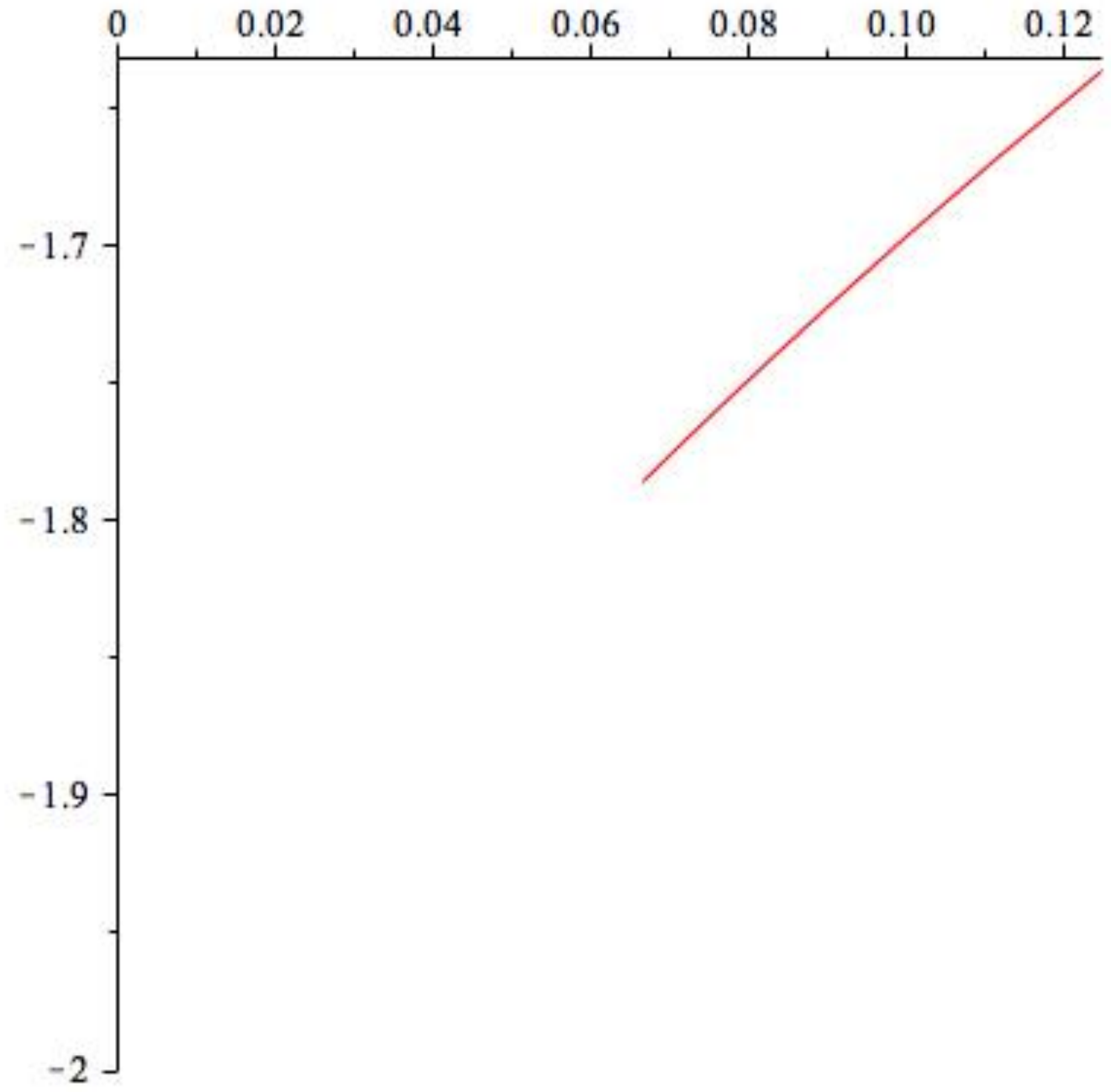}
\end{center}
\caption{Local gradient of $\log(B^{(1)}_h(x_{\rm c}))$ as a function of $\log {h}$.}
\label{fig:Bh-gradient}
\end{figure}

In \cite{BeatonGuttmannJensen12} the values of  $A_h(x_c)$ and $B_h(x_c)$ were calculated for SAWs in a strip of width $h$ on the square lattice for $h < 16.$ The SLE result that $B_h(x_{\rm c}) \sim b_1 \cdot h^{-1/4}$ is independent of lattice, so from the square lattice data it should be possible to estimate the asymptotics in more detail. If, for the square lattice,
\begin{equation}
 B_h(x_{\rm c}) \sim \frac{b_1}{h^{1/4}} + \frac{b_2}{h^{\beta_2}} \,,
\end{equation}
then defining
\begin{equation}
 {\tilde B}_h(x_{\rm c}) =h^{1/4}\cdot B_h(x_{\rm c}) \,,
\end{equation}
we have
\begin{equation}
 B^{(1)}_h(x_{\rm c}) = {\tilde B}_h(x_{\rm c}) - {\tilde B}_{h-1}(x_{\rm c}) \sim \frac{-b_2}{\beta_2 h^{\beta_2+3/4}} \,.
\end{equation}
A plot of $\log(B^{(1)}_h(x_{\rm c}))$ against $\log {h}$ looks linear, and plotting the local gradient produces Figure~\ref{fig:Bh-gradient},
which is rather clearly extrapolating to a value around $-2$. From this we conclude that $\beta_2 = 5/4.$

So it appears that $B_h(x_{\rm c})$ decays as $1/h^{k+1/4}$ for $k=0,1,\ldots.$ We studied this in greater detail, and found
\begin{equation}
 B_h(x_{\rm c}) \sim \frac{1.00180}{h^{1/4}}-\frac{0.514}{h^{5/4}}+\frac{0.69}{h^{9/4}} \,,
\end{equation}
where errors in the constants are expected to be confined to the last quoted digit.
A similar analysis of $\cos \left ( \frac{3\pi}{8} \right )A_{h}(x_{\rm c})$ gives
\begin{equation}
 \cos \left ( \frac{3\pi}{8} \right )A_{h}(x_{\rm c})  \sim 1.024966- \frac{1.00180}{h^{1/4}}+\frac{0.514}{h^{5/4}}-\frac{0.76}{h^{9/4}} \,.
\end{equation}
Note that, upon addition, the two terms of order $h^{-1/4}$ and $h^{-5/4}$ cancel, though not the term of order $h^{-9/4}.$ This clearly illustrates the approximate nature of the ``identity'' (\ref{DCS-identity}) in the case of the square lattice. As we have seen, for the honeycomb lattice all terms in the asymptotics must cancel, whereas for the square lattice the two leading-order terms cancel but the third-leading term does not.

So we conjecture that for the square lattice,
\begin{equation}
 \cos \left ( \frac{3\pi}{8} \right )  \cdot A_h(x_{\rm c}) + B_h(x_{\rm c}) = c_1+\frac{c_2}{h^{9/4}} +o(h^{-9/4}) \,,
\end{equation}
since all terms in the $h$-dependent asymptotics of the individual generating functions of order $1/h^\theta$ cancel upon addition,  for $\theta < 2.$ From the above analysis of the individual generating functions we have that $c_1 \approx 1.024966$ and $c_2 \approx -0.07.$

In \cite{BeatonGuttmannJensen12} it was suggested that for SAWs in strips of width $h$,  it appeared that
$$\cos \left ( \frac{3\pi}{8} \right )  \cdot A_h(x_{\rm c}) + B_h(x_{\rm c}) = c_1+c_2/h^2 +o(h^{-2}) \,.$$
$c_1$ was estimated to be $c_1 \approx 1.024966,$ and the correction term exponent was conjectured to be around $2,$ but as the more informed analysis above shows, it is actually $9/4.$

In \cite{BeatonGuttmannJensen12} the values of $A_h(x_{\rm c})$ and $ B_h(x_{\rm c})$ were estimated by evaluating the generating functions---generated to 1000 terms---at the best estimate of $x_{\rm c}.$ Numerical experiments since then show that this may lead to some loss of accuracy as the strip width $h$ increases, as the radius of convergence moves closer to $x_{\rm c}$ as $h$ increases. So  here we estimate instead the value of $A_h(x_{\rm c})$ and $ B_h(x_{\rm c})$ by constructing Pad\'e approximants from the 1000 terms we have of the ogfs. We find good convergence at all widths (up to 16) with just 500 terms---that is to say, calculating the $[250,250]$ approximants gives 26 or 27 digit accuracy.

So based on this more accurate evaluation, and a better understanding of the asymptotics,  we have re-analysed the data. 
First, we evaluated
\begin{equation}
 f(h)=\cos \left ( \frac{3\pi}{8} \right ) A_h(x_{\rm c})+B_h(x_{\rm c}) \,,
\end{equation}
using for $x_{\rm c}$ the value (\ref{the-conjecture}). This leads to the data shown in Table~\ref{table1}.
Then we estimated the asymptotics of $f(h)$ directly, and found $$f(h)=1.024966-\frac{0.09}{h^{9/4}} \,.$$
This is consistent with the results above found for the two generating functions independently.

\begin{table}[htp]
\begin{center}
\begin{tabular}{|r|l|}
\hline
$h$ & $f(h) =\cos \left ( \frac{3\pi}{8} \right ) A_h(x_{\rm c})+B_h(x_{\rm c})$ \\
                     \hline
 1&                    1.02219272918820979664488870725378\\
   2&                  1.02334333088068448056310636457412\\
   3&                  1.02389842811365227525953579448892\\
   4&                  1.02421016363655290027640666164740\\
  5&                   1.02440273931572277711772000092385\\
 6&                    1.02453002555478382232391817150095\\
  7&                   1.02461854312010049603949089063169\\
 8&                    1.02468258891795855900214615905381\\
  9&                   1.02473042540555691116124731802323\\
  10&                   1.02476709890685031987372860087368\\
 11&                    1.02479583338438063405244906222760\\
  12&                   1.02481876703019201367883647551132\\
  13&                   1.02483736342864416741456527040520\\
  14&                   1.02485265184376158282518177725271\\
  15&                   1.02486537324197170618110312415885\\
 %  16&                  1.02487607207157647286997091767877\\
\hline
\end{tabular}
\end{center}
\caption{$f(h)=\cos \left ( \frac{3\pi}{8} \right )  A_h(x_{\rm c})+B_h(x_{\rm c})$ values for strips of width $h$.}
%\textcolor{red}{\bf Removed the strange factor of c from the second column caption.}}
\label{table1}
\end{table}

To use these results to estimate the critical point, $x_{\rm c}$, define $$f_h(x)=cA_h(x)+B_h(x),$$ where $c= \cos \left ( \frac{3\pi}{8} \right )$.
Then $f_h(x) \sim f_h(x_{\rm c})+(x-x_{\rm c})f'_h(x_{\rm c}).$ From our data we can also investigate the $h$ dependence of $f'_h(x_{\rm c}).$ Writing $f'_h(x_{\rm c}) \sim d_0+d_1h^\alpha,$ a simple ratio plot of the ratios $f'_h(x_{\rm c})/f'_{h-1}(x_{\rm c})$ should have gradient $\alpha$ when plotted against $1/h.$ In this way we estimated $\alpha = 1.095 \pm 0.005.$ We also have shown above that $f_h(x_{\rm c}) \sim c_0+c_1h^\beta,$
where $\beta=-9/4.$ Now solving $$f_h(x)=f_{h+1}(x),$$ gives solution $x=x_h.$ From the preceding, a little algebra yields $$x_h=x_{\rm c}-\frac{f_h(x_{\rm c})-f_{h+1}(x_{\rm c}) }{f'_h(x_{\rm c})-f'_{h+1}(x_{\rm c})},$$ so that
\begin{equation}
 x_h\sim x_{\rm c}+ {\rm const} \times h^{\beta -\alpha} =x_{\rm c}+ {\rm const} \times h^{-3.345} \,.
 \label{eq16}
\end{equation}
So this calculation tells us how we should extrapolate the sequence $\{x_h\}.$
 
Solving $\cos \left ( \frac{3\pi}{8} \right ) A_h(x)+B_h(x) = \cos \left ( \frac{3\pi}{8} \right )  A_{h+1}(x)+B_{h+1}(x),$ for $h=1,\ldots,14,$  gives the sequence
of estimates of $x_h,$ shown in Table \ref{xh}.

\begin{table}[htp]
\caption{$x_{\rm c}(h)$ values from strips of width $h$ and $h+1.$ }
\begin{center}
\begin{tabular}{|r|l|}
\hline
$h$ & $x_{\rm c}(h)$ \\
                     \hline
   1&                  0.378849252443011782370230729987601\\
     2&                0.378956388842518611129966996155422\\
   3&                  0.378999327220638659189584698621997\\
      4&              0.379020021122802690310729735547924\\
5&                     0.379031210267367973140729069860968\\
       6&              0.379037779036960947578360165143159\\
     7&                0.379041884134152407919456202260939\\
   8&                 0.379044579076221113573987098156221\\
 9&                    0.379046420146306777562797592471999\\
      10&               0.379047719933799254093039766931429\\
 11&                    0.379048663277895734792965515600497\\
    12&                 0.379049364230541388319754913510459\\
  13&                   0.379049895752829776131162650396880\\
  14&                   0.379050305989827925108903738816740\\
 15&                    0.379050627578933937642251833271687\\
 %16 &                  0.379050883174069115180825963236831\\
\hline
\end{tabular}
\end{center}
\label{xh}
\end{table}

As suggested by the above analysis, we extrapolated this sequence against $h^{-3.345}.$ Call these extrapolants $x_h^{(1)}.$ We then extrapolated $x_h^{(1)}$ against $h^{-4.345},$ giving a new sequence $x_h^{(2)}.$ We iterated this process, and then extrapolated the sequence $\{x_{15}^{(n)}\}$ against $1/n.$ In this way we arrived at the estimate
\begin{equation}
  x_{\rm c} = 0.379052277750 \pm 0.0000000005 \,.
\end{equation}
One sees that this agrees with the value (\ref{the-conjecture}) of the mnemonic polynomial, with uncertainty confined to the 11th digit. So this method is less powerful than that based on series analysis of the polygon series, but in fairness it must be said that the series analysis approach has had significantly greater computational resources devoted to it. With strips of greater width, it is possible one could achieve comparable precision. However convergence of the estimates is not particularly rapid, with each iterate only improving things by a factor $1/h.$ 

By contrast, the new method we discuss in the next section provides an improvement by a factor $1/h^2$ with each iteration. Consequently, the new method described below is substantially more powerful, so there seems little point in pursuing this approach to estimate $x_{\rm c}$. However this analysis has yielded considerable insight into the identity of Duminil-Copin and Smirnov, and the behaviour of associated generating functions for the hexagonal and square lattices, and so is valuable from that perspective too.

\section{Topological transfer matrix method}

The method that we now describe originates from a graph polynomial construction \cite{JacobsenScullard12,JacobsenScullard13,Jacobsen14,Scullard11-2} which can be used
to determine the critical manifold of the $q$-state Potts model on regular two-dimensional lattices, not only in the usual case (which we shall need here) with homogeneous
couplings, but also in a more general setup with periodic inhomogeneities \cite{JacobsenScullard12} or even quenched bond randomness \cite{OhzekiJacobsen15}.

The original graph polynomial $P_B(q,v)$ was defined in \cite{JacobsenScullard12} as a topologically weighted partition function on a finite piece $B$, called a {\em basis}, of
the lattice under consideration. It has two crucial properties:
\begin{itemize}
\item[(1)] If the model is exactly solvable, then $P_B(q,v)$ has a root exactly at the critical temperature, $v = v_{\rm c}$.
\item[(2)] If it is not, then for each size $n$ of the basis, $P_B(q,v)$ has a real root $v_{\rm c}(n)$ that converges rapidly to $v_{\rm c}$ as $n \to \infty$.
\end{itemize}
Since the self-avoiding walk model on the square lattice is not believed to be exactly solvable, we are here interested mainly in the second property.

A computationally efficient means of computing $P_B(q,v)$ is to use a transfer matrix construction. The initial scheme \cite{JacobsenScullard13} was considerably
improved in \cite{Jacobsen14} by noticing that the boundary conditions on $B$ are essentially doubly periodic, so that the transfer matrix can be simplified by
using results from the representation theory of the periodic Temperley-Lieb algebra. A further improvement was achieved in \cite{Jacobsen15} by realising that
if $B$ is taken infinite in one direction---concretely, as a semi-infinite cylinder of circumference $n$---then the determination of $v_{\rm c}(n)$ amounts to
equating the leading eigenvalues in two topologically distinct sectors. This has the advantage of dramatically diminishing the computational complexity (again!), since
the transfer matrix needed to deal with the $n \times \infty$ basis has dimension $4^n$, as opposed to the $16^n$ needed in \cite{JacobsenScullard13}
to deal with a finite $n \times n$ basis. As a matter of fact, the transfer matrix construction of \cite{Jacobsen15} turns out to be not only more efficient, but also
much simpler than that of \cite{Jacobsen14}, since only one time slice is needed (instead of two).

It was also realised in \cite{Jacobsen15} that the topological transfer matrix method can be generalised from the $q$-state Potts to the O($N$) model, while
maintaining the two crucial properties cited above. In particular, it applies to self-avoiding walk models, upon taking the usual $N \to 0$ limit. Property
(1) is exemplified by the fact that the method determines the growth constant $\mu_H$ of SAWs on the hexagonal lattice \cite{Nienhuis82,DuminilSmirnov12}
exactly for any $n$, with no finite-size dependence. This property is thus very much reminiscent of the identity (\ref{DCS-identity}), although it is presently
not as well understood from the mathematical perspective.

We shall need here instead property (2). It is illustrated by the fact that the effective critical point $x_{\rm c}(n)$ for square-lattice SAWs, as determined by the topological
transfer matrix method, does depend on $n$. It was determined in \cite[Table~5]{Jacobsen15} for $2 \le n \le 19$, and although the data analysis was left out of that
paper, it was manifest that $x_{\rm c}(n)$ converges to the true critical point $x_{\rm c}$ very fast. Below we shall describe a parallel implementation of the algorithm
of \cite{Jacobsen15} which will allow us to determine two more data points, $n=20$ and $n=21$. We then analyse this data carefully, in order to extract the
best possible value of $x_{\rm c}$. But before doing so, we review the necessary ingredients of the method (see \cite{Jacobsen15} for more details), paying special
attention to issues of computational complexity.

\subsection{Methodology}

The method builds on the evaluation of the largest eigenvalue of the SAW transfer matrix $T$ in two distinct topological sectors. The setup is almost identical to that of
section~\ref{sec:series}, and in particular Figure~\ref{fig:saptransfer} and the surrounding discussion can be taken over almost unchanged. More specifically, the coding of
the connectivity states, the updating rules, and the (sparse matrix) factorisation of $T$---obtained by building up the lattice one vertex at a time---are essentially the
same. We therefore outline only the differences, referring the reader to \cite{Jacobsen15} for more details.

The most important difference is that we do not work on a strip, but on a cylinder with periodic boundary conditions in the $n$-direction. This means that the perfectly
balanced parenthesis system made of openings and closings of loop strands along the boundary line (coded as `1's and `2's) should be read cyclically. However, this
does not change the fact that the number of configurations is still related to the Motzkin numbers. The periodic boundary conditions amount to treating the boundary line
as a periodic object, called ``auxiliary space'' in \cite{Jacobsen15}, borrowing the terminology of quantum integrable systems. In the same vein, the operator that
locally adds one vertex to the lattice (at the position $i$ of the kink) is denoted $\check{\sf R}_i$. It contains seven distinct diagrams \cite[eqs.~(52) and (68)]{Jacobsen15}.

Another, more minor, difference is that we wish to determine the largest eigenvalue of $T$, rather than using it to perform exact enumerations.
Therefore we do not employ the concept of pruning. Moreover, each entry of $T$ is a real number (the Boltzmann weight) rather than a formal power
series with integer coefficients, as would be required for enumerative purposes. This considerably reduces the amount of time and storage needed to deal with one
entry in $T$, although admittedly we lose the reduction of the state space which would result from the pruning.

To diagonalise $T$, the simplest option is to apply the power method. This means that we apply repeatedly $T$ to a well-chosen initial vector, observing the
growth of norm $\Lambda$ after each application. We stop the iteration when $\Lambda$ has converged to the required numerical precision (here chosen as
40-digit precision) and identify the result of the last iteration with the sought eigenvalue.

The diagonalisation is performed in two distinct sectors, corresponding to two different choices of the initial vector. In sector 0, the transfer matrix is denoted $T^{(0)}$,
and there is a weight of
\begin{equation}
  N_{\rm wind} = -\sqrt{2-N} = -\sqrt{2}
\end{equation}
for each loop wrapping the periodic direction of the cylinder. Recall that we have taken $N \to 0$, so that loops which are homotopic to a point are forbidden.
The initial vector can be taken simply as the empty state (coded as `0's for all $i$).
The fact that $N_{\rm wind} < 0$ will adversely affect the efficiency of the power method, so that sector 0 is the computationally most demanding. Below we
give concrete examples of the number of multiplications needed to obtain convergence.

The other sector we shall need is called sector 1, and its transfer matrix is denoted $T^{(1)}$. It corresponds to having an open loop segment (a SAW actually)
running along the transfer direction. This can be implemented as one unmarked occupied point in the otherwise perfectly balanced parenthesis system.
The updating rules are otherwise unaltered by this modification. The initial state can be taken as the state with the SAW residing at the leftmost point, and otherwise empty.

Let $\Lambda^{(0)}$ and $\Lambda^{(1)}$ denote the largest eigenvalues of $T^{(0)}$ and $T^{(1)}$ respectively, for a given size $n$. The principle of
the topological transfer matrix method is to find the value $x_{\rm c}(n)$ of the monomer fugacity for which
\begin{equation}
 \Lambda^{(0)} = \Lambda^{(1)} \,.
\end{equation}
To this end we adjust $x$ in a second-order Householder scheme \cite[section 5]{Jacobsen15} which requires three evaluations of either eigenvalue for each
iteration. For large values of $n$, the preceeding data permits us to predict the initial value for $x_{\rm c}(n)$ to such a precision that only one or two Householder
iterations are needed to attain the desired precision.

\subsection{Parallel implementation}

We adapt for our purposes the parallel algorithm used by Jensen \cite{Jensen03} to enumerate self-avoiding polygons. To remain consistent with \cite{Jacobsen15},
we take the transfer direction to be upwards (note by contrast that Figure~\ref{fig:saptransfer} transfers towards the right). 
As described above, as we build up the lattice with the transfer matrix, the top row consists of edges that are either the right or left ends of a loop, the end of a bridge, or are ``empty'' meaning they are not part of any loop or bridge. If we term all loop and bridge ends as ``occupied'', the foundation of the parallel implementation is the observation that the local $\check{\sf R}_i$ operators of the transfer matrix cannot affect the occupied or empty status, which Jensen calls the occupation pattern, of distant loop ends.

Two goals of any parallel algorithm should be to minimise communication between processors and balance the workload so that processors are not sitting idle waiting for others to complete. The Jensen algorithm \cite{Jensen03} accomplishes both of these goals in an elegant manner. The state vector is broken up and distributed to the different tasks, which then each work on their own piece. The action of $\check{\sf R}_i$ at site $i$ produces a new state, whose weight must be modified and stored. If applying one of these operators to a state on a given processor produces a state not owned by that processor then communication is going to be needed to put the weight in the appropriate place. We would like to minimise this occurrence or eliminate it altogether. This is addressed in Jensen's algorithm by dividing the boundary line into two segments. If there are $n$ loop ends, then the first half might contain the first $\lfloor n/2 \rfloor$, and the second half all the rest. When the transfer matrix is operating in the first half, states are grouped according to their occupation pattern on the second half; all states with the same occupation pattern are placed on the same processor. In this way, one is assured that an operation of $\check{\sf R}_i$ by a given processor will only produce a state with a weight that is already on that processor. This eliminates the need for communication up until the segment boundary is reached by the kink edge. At that point, the states must be regrouped according to their occupation pattern on the first half of the boundary line. This implies that communication is necessary to reorganise the states among the processors. Likewise, upon reaching the end of a row this must be done again so that the states are once again grouped according to the pattern on the second half. As occupation patterns are essentially just binary numbers their mapping to integers is obvious.

Now, for a given occupation pattern on a given half of the boundary line there corresponds a number of states. But these are not distributed equally for every occupation pattern, with some patterns being far more frequent than others. One is then faced with the problem of ensuring that each processor has approximately an equal number of states. This is accomplished by creating a frequency table of occupation patterns. This is then sorted and the states are distributed according to an algorithm which is thoroughly described in \cite{Jensen03} and to which we therefore refer the reader for details.

Here, we implement Jensen's algorithm with only a few minor changes. First, we divide a row into multiple segments, not just two. The reason is that this affords more freedom in choosing the number of processors. Using more segments naturally allows us to divide the states more finely into occupation patterns and thus to use more processors with fewer states on each. The tradeoff is that we must do the reorganisation step more frequently as there are now more segment boundaries to cross.

We also divide the calculation into three phases. The first is the startup phase, when relatively few states have yet been generated. Here, many occupation patterns are not represented and therefore have frequency zero. At this stage we do not bother to sort the frequency table, and occupation patterns with non-zero frequency are simply assigned in sequential order to the processors. This results in a poorly balanced system but it avoids the wasteful step of sorting lists where most entries are zero. The second phase is when all occupation patterns are represented in the system by at least one state, but new states are still being generated. During this phase, we use Jensen's distribution algorithm to balance the load. The third phase is where all the states are present in the system, which is identified when the number of states is unchanged upon completing a full cycle. At this point, we no longer need to do the sorting and distribution steps because the results will be no different from the last time the state ownership was computed for that segment, so we just save these results in tables and refer to them for all subsequent reorganisations.

The final modification we make is that we perform the state reorganisation steps after inserting and removing the auxiliary space \cite[section 3.5.5]{Jacobsen14}.

\subsection{Performance and resource allocation}

We run on Lawrence Livermore National Laboratory's Cab and Vulcan supercomputers. Each processor on Cab is a 2.6 GHz Intel Xeon E5-2670 with 2GB of RAM, while Vulcan is built for massively parallel jobs with lower individual processor specs, 1.6GHz Power PC A2 with 1GB each, but one can access many more processors per job. The parallel algorithm was primarily used to compute $x_{\rm c}$ for $n=20$ and $21$, although we also used it to complete the final Householder iterations for $n=19$.%
\footnote{Namely, in \cite[Table 5]{Jacobsen15} the $n=19$ result was only given to 22-digit precision, instead of the usual 40 digits.}

For $n=20$, we computed sector 0 on Cab, using $768$ processors and we divided the top row into four segments. In four hours, we were able to compute about 57 iterations of the power method (i.e., 57 multiplications by $T^{(0)}$). Starting with the initial vector described above, we found that we needed about $831$ power method iterations to obtain convergence of the eigenvalue. However, when we run it again with an updated $x_{\rm c}(n)$, we use the final vector obtained from the previous value and this reduces the number of iterations needed to $533$ so that the first run is the most expensive. We handled sector 1 on Vulcan, and there we ran with $1024$ processors. Sector 1 being the easier of the two to converge, we needed about $296$ power method iterations starting from the initial vector but only $96$ when we used the vector from the previous value of $x_{\rm c}(n)$.

We performed the $n=21$ calculation on Cab for both sectors. Here, we divided the row up into seven segments and used 2400 processors. This resulted in approximately $1.8 \times 10^6$ states on each processor and we were able to complete about 18 power method iterations in an hour. For sector 0, the maximum number of iterations needed was 871 and for sector 1 we needed 310.

\subsection{Computational complexity}

Neglecting for simplicity the cost of the parallelisation scheme, the consumption of time and memory mainly depends on the dimension of the transfer matrices
$T^{(0)}$ and $T^{(1)}$. Exact expressions for these have been given in \cite[eq.~(66)]{Jacobsen15}. Either dimension has the same asymptotic growth,
\begin{equation}
 {\rm dim}\left( T^{(0)} \right) \sim {\rm dim}\left( T^{(1)} \right) \sim \frac12 \left( \frac{3}{\pi n} \right)^{1/2} 3^n \,.
\end{equation}

The number of transfer matrix multiplications (i.e., power method iterations) needed to achieve convergence in each sector is reported in Table~\ref{tab:numiter}.
It is easily seen that these numbers are proportional to $n$, up to numerical rounding effects. In addition, 
it should not be forgotten that each $T$ consists of $n$ factors $\check{\sf R}_i$. We conclude the memory and time consumption grow like $n^{-1/2} 3^n$
and $n^{3/2} 3^n$, respectively.

\begin{table}[htp]
\begin{center}
\begin{tabular}{|r|r|r|}
\hline
$n$ & Sector 0 & Sector 1 \\ \hline
16 & 679 & 239 \\
17 & 721 & 253 \\
18 & 755 & 267 \\
19 & 790 & 281 \\
20 & 831 & 296 \\
21 & 871 & 310 \\ \hline
\end{tabular}
\end{center}
\caption{Number of iterations of $T^{(k)}$ needed for 40-digit numerical convergence in sector $k=0,1$ for various sizes $n$.}
\label{tab:numiter}
\end{table}

The series method, which has been reviewed in section~\ref{sec:series}, relies on a very similar transfer matrix construction, where again the dimension grows like $3^n$.
One cannot however directly compare their time complexities, for the simple reason that they do not compute the same quantities. For the sake of the matter,
let us adopt here the simplistic view that we only wish to know which method can determine $x_{\rm c}$ to the highest numerical precision for a given computational effort.
Obviously, then, the answer will depend on two factors:
\begin{itemize}
 \item[(i)] How large sizes ($w$ or $n$) can be attained with a given amount of resources?
 \item[(ii)] What is the rate of convergence of $x_{\rm c}$ as a function of that size?
\end{itemize}

To roughly appreciate the first criterion, we can provide a back-of-the-envelope estimate of the resources that have been spent by either of the two methods at this stage.
For the series method, Clisby and Jensen \cite{ClisbyJensen12} have attained $w_{\rm max} = 32$, corresponding to a maximal SAP perimeter of
$4 w_{\rm max} + 2 = 130$ steps. Given the advantages of pruning, this corresponds to a transfer matrix dimension of the order $2.06^{32} \approx 1.11 \times 10^{10}$.
For the topological transfer matrix method, we have here attained $n_{\rm max} = 21$. This corresponds to a dimension of the order $3^{21} \approx 1.05 \times 10^{10}$.
While these two numbers are of the same order of magnitude, this quick comparison did not take into account that the ``entries'' of the transfer matrix are of different nature.
For the series method they are polynomials with large integer coefficients, while for the topological method they are high-precision real numbers.

We cautiously conclude that the two methods have consumed a similar amount of resources. The outcome of the comparison will therefore largely depend
on the second criterion, the data analysis, to which we turn next.

\subsection{Results and data analysis}

Our results for $x_{\rm c}(n)$ are gathered in Table~\ref{tab:xc}. We have analysed them by going through the same reasoning that
was applied in \cite[section 7.1]{Jacobsen15} for site percolation thresholds on the square lattice.

\begin{table}[htp]
\begin{center}
\begin{tabular}{|r|l|}
\hline
$n$ & $x_{\rm c}(n)$ \\ \hline
 2  & 0.3832870437289217825415444959209990643484 \\
 3  & 0.3800152822923947541103727449094743052839 \\
 4  & 0.3793419092420152604076859124268482909456 \\ 
 5  & 0.3791615386298805591124869699564102732536 \\
 6  & 0.3791017465104568577033096312174651793134 \\
 7  & 0.3790779263723816763857349117326710080035 \\
 8  & 0.3790669419366251682820022783255996752011 \\
 9  & 0.3790612863965732376129739341339159714858 \\
10 & 0.3790581237478262657302859193323348704028 \\
11 & 0.3790562392439348634338963536547147709970 \\
12 & 0.3790550583590770828697993099179842253186 \\
13 & 0.3790542873705249946097446478792002255473 \\
14 & 0.3790537664746062070854620937409756594548 \\
15 & 0.3790534041836437725305420784870138649786 \\
16 & 0.3790531458388626510867578645132848654379 \\
17 & 0.3790529575762840825464391257666224019613 \\
18 & 0.3790528177462476184521578621884148510125 \\
19 & 0.3790527121228867470584088247673298012074 \\
20 & 0.3790526311295114746816952601274560023696 \\
21 & 0.3790525681789990003293315280416877312746 \\ \hline
\end{tabular}
\end{center}
\caption{Results for $x_{\rm c}(n)$. The data with $n \le 18$ and the first 22 digits of $n=19$ already appeared in \cite{Jacobsen15}.}
\label{tab:xc}
\end{table}

In a first step, we assume a standard power law scaling of the form
\begin{equation}
 x_{\rm c}(n) = x_{\rm c} + \sum_{k=1}^\infty \frac{A_k}{n^{\Delta_k}} \,.
 \label{FSS}
\end{equation}
Taking discrete logarithmic derivatives of $x_{\rm c}(n) - x_{\rm c}$, using initially the value (\ref{the-conjecture}), and
fitting the finite-size estimates to polynomials in $1/n$, we establish that $\Delta_1 = 4.000\,000\, (1)$. Assuming now
that $\Delta_1 = 4$ exactly and subtracting off this term, we obtain in the same way $\Delta_2 = 6.000\, (4)$. So we
may safely conclude that $\Delta_2 = 6$.

Repeating the scheme to get $\Delta_3$, we start seeing signs that maybe (\ref{the-conjecture}) is not exact after all.
Adjusting it slightly on the 12th decimal, we get results compatible with $\Delta_3 \approx 8$. We shall henceforth assume
that
\begin{equation}
 \Delta_k = 2(k+1) \,, \quad \mbox{for any } k \ge 1
 \label{FSS1}
\end{equation}
and use this as an input for the second step of the analysis.

Let $n_{\rm max}=21$ denote the largest size for which $x_{\rm c}(n)$ is known. We first form a series of estimators $x_{M,L}$
from $x_{\rm c}(n)$, by truncating the scaling form (\ref{FSS})--(\ref{FSS1}) at the $1/n^M$ term and using the data $x_{\rm c}(n)$
up to a maximum size of $n=L$. Stated otherwise, we find the unique solution of the linear system
\begin{equation}
 x_{M,L} + \left( \frac{A_1}{n^4} + \frac{A_2}{n^6} + \cdots + \frac{A_{M/2-1}}{n^M} \right) = x_{\rm c}(n) \,,
\end{equation}
with $n=L+1-M/2,\ldots,L-1,L$. Next, for fixed $M$, we form another series of estimators $x_{M}^{(n_0)}$ from $x_{M,L}$, by
fitting the latter to the residual dependence predicted by (\ref{FSS})--(\ref{FSS1}), but eliminating from the fit the first $n_0$ 
possible values of $L$. That is, we find the unique solution of the linear system
\begin{equation}
 x_{M}^{(n_0)} + \left( \frac{B_1}{n^{M+2}} + \frac{B_2}{n^{M+4}} + \cdots + \frac{B_{n_{\rm max}-n_0-1-M/2}}{n^{2(n_{\rm max}-n_0-1)}} \right) = x_{M,L} \,.
 \label{2ndfit}
\end{equation}
This fit uses $n_{\rm max}-n_0-M/2$ different values of $L$, ranging from $1+M/2+n_0$ up to $n_{\rm max}$.

For a fixed order $M$, we now study the variation of $x_M^{(n_0)}$ with $n_0$. When too few data points at small sizes have been
eliminated (i.e., $n_0$ is taken too small) we cannot expect $x_M^{(n_0)}$ to approximate $x_{\rm c}$ very well, since (\ref{FSS})
only holds asymptotically. But, on the other hand, when $n_0$ is taken too large, the fit (\ref{2ndfit}) will not have enough terms and the 
precision will again deteriorate. Therefore we expect an optimum in between these extremes.

In practice we observe that $x_M^{(n_0)}$ is almost constant up to a value $n_0^\star \simeq 7$ (that depends only very slightly on $M$),
whereas for $n_0 > n_0^\star$ it drops off abruptly. This provides compelling evidence that $n_0^\star$ is optimal and that
$x_M^{(n_0^\star)}$ is an accurate estimate for $x_{\rm c}$. Indeed, we observe that $x_M^{(n_0^\star)}$ is almost independent
of the order $M$ of the approximant, provided of course that the latter is neither too small, nor too large. Comparing the values
for $8 \le M \le 16$ we obtain a final value and error bar
\begin{equation}
 x_{\rm c} = 0.379052277755161\, (5) \,.
\end{equation}

We have validated the method by reiterating the whole procedure for $n_{\rm max}=20$ and $n_{\rm max}=19$, verifying that 
we indeed get compatible (and of course more accurate) results as the number of data points is increased.

\section{Conclusion}
Our principal result is a significantly more precise estimate of the growth constant for the square lattice SAW. It is
$$\mu = 2.63815853032790\, (3).$$
We review the Finite-Lattice Method which has been the most successful method until recently for generating series expansions allowing estimates of $\mu$ to be made. We also review a more recent method based on extending a quantity which is a lattice invariant on the hexagonal lattice, discovered by Duminil-Copin and Smirnov \cite{DuminilSmirnov12}, to the square lattice, where it is not an invariant. But by establishing its convergence properties, we are able to find quite precise estimates of the growth constant for other lattices. The third method, which we call the Topological Transfer Matrix (TTM) method,  originates from a graph polynomial construction \cite{JacobsenScullard12,JacobsenScullard13,Jacobsen14} which has been used previously
to determine the critical manifold of the $q$-state Potts model on regular two-dimensional lattices.

The original graph polynomial $P_B(q,v)$ was defined in \cite{JacobsenScullard12} as a topologically weighted partition function on a finite piece $B$, called a {\em basis}, of the lattice under consideration.  In \cite{Jacobsen15} the topological transfer matrix method was generalised from the $q$-state Potts to the O($N$) model. Here we report the results of devoting considerable computing resources to this problem, in order to obtain a significantly more precise estimate of the growth constant $\mu.$ 

If further computational resources are devoted to this or allied problems, it appears that the TTM method is the appropriate choice, as asymptotically the estimates converge as $1/L^4,$ with higher order corrections converging as $1/L^2.$ For the method based on an adapted identity of Duminil-Copin and Smirnov, convergence is a little slower---being $1/L^{3.345}$ according to (\ref{eq16})---and higher order corrections converge only linearly, compared to quadratically for the TTM. For the traditional method of series analysis we cannot make such a direct comparison, as estimates of the growth constant are obtained from differential approximants, and an analysis of the rate of convergence with series length has not been made. However, using comparable resources the TTM method provides an estimate of the growth constant with uncertainty in the 15th digit, compared to series analysis, in which the uncertainty is in the 12th digit. This implies that the TTM method is the most rapidly convergent, and thus justifies the use of further computational resources.

In the future it should be possible to substantially increase the precision of the estimate of the growth constant for SAWs on other unsolved lattices, such as the triangular lattice,
by use of the TTM method.

\section*{Acknowledgements}
JLJ is grateful for the hospitality of the Centre of Excellence for Mathematics and Statistics of Complex Systems (Melbourne University) where part of this
work was accomplished. He also acknowledges the support of the Institut Universitaire de France, and of the European Research Council through the Advanced Grant NuQFT.
The work of CRS was performed under the auspices of the U.S.\ Department of Energy at the Lawrence Livermore National Laboratory under Contract No.\ DE-AC52-07NA27344. AJG acknowledges the support of the Australian Research Council through grant DP120100939. We thank Mireille Bousquet-M\'elou for the provision of a figure.

\section*{References}
\bibliographystyle{unsrt}
\bibliography{JSG15}

\end{document}